\documentclass{elsart}
\makeatletter
\def\fps@figure{p}
\def\fps@table{p}
\makeatother
\usepackage{graphicx}
\newcommand\UT{University of Tokyo}
\newcommand\hongo{7-3-1~Hongo, Bunkyo-ku, Tokyo~113-0033, Japan}
\newcommand\gagg{g_{a\gamma\gamma}}

\begin{document}
\begin{frontmatter}
%
%
\title{Search for solar axions with mass around 1 eV
 using coherent conversion of axions into photons}

%
%
\author[ICEPP,RESCEU]{Y. Inoue},
\author[PHYS]{Y. Akimoto},
\author[PHYS]{R. Ohta},
\author[PHYS]{T. Mizumoto},
\author[KEK,RESCEU]{A. Yamamoto},
\author[PHYS,RESCEU]{M. Minowa\corauthref{AUTH}}
\corauth[AUTH]{Corresponding author}
\ead{minowa@phys.s.u-tokyo.ac.jp}
\address[PHYS]{Department of Physics, School of Science, \UT, \hongo}
\address[ICEPP]{International Center for Elementary Particle Physics,
  \UT, \hongo}
\address[KEK]{High Energy Accelerator Research Organization (KEK),
  1-1~Oho, Tsukuba, Ibaraki~305-0801, Japan}
\address[RESCEU]{Research Center for the Early Universe (RESCEU),
School of Science, \UT, \hongo}
%
%
\begin{abstract}
A search for solar axions has been performed
using an axion helioscope
which is equipped with
a 2.3-m long 4\,T superconducting magnet,
a gas container to hold dispersion-matching gas,
PIN-photodiode X-ray detectors,
and a telescope mount mechanism to track the sun.
A mass region around
$m_a = 1{\rm\,eV}$
was newly explored.
From the absence of any evidence,
analysis sets a limit on axion-photon
coupling constant to be
$\gagg<\mbox{5.6--13.4}\times10^{-10}\rm GeV^{-1}$
for the axion mass of
$0.84<m_a<1.00\rm\,eV$
at 95\% confidence level.
It is the first result to search for the axion in the
$\gagg$-$m_a$ parameter region of the preferred axion models
with a magnetic helioscope.
\end{abstract}

%
%
\end{frontmatter}

%
%
\section{Introduction}
Quantum chromodynamics (QCD) is the theory of the strong interactions.
Although QCD has proven remarkably successful,
there is a blemish called the strong $CP$ problem.
The strong $CP$ problem is that the effective Lagrangian
of QCD has $CP$ violating term but it is not observed,
i.e.,
the experimental value of neutron electric dipole moment is
smaller than expected
by many orders of magnitude.
Peccei and Quinn proposed an attractive solution to solve
this problem
\cite{axion-bible1,axion-bible2,axion-bible3,axion-bible4,axion-bible5}.
They introduced a new global $U(1)$ symmetry,
Peccei--Quinn (PQ) symmetry.
When PQ symmetry is spontaneously broken,
a new effective term arises in QCD Lagrangian
which cancels the $CP$ violation term.
The solution also predicts a new pseudo Nambu--Goldstone
boson, axion.
The expected behavior of an axion is characterized
mostly by the scaling factor of the PQ symmetry breaking, $f_a$,
and so its mass, $m_a$, which is directly related to $f_a$ by
$m_a=6\times10^{15} [\mathrm{eV}^2]/f_a$.

Axions are expected to be produced in stellar core through their coupling to photons with energies
of order keV.
Especially,
the sun can be a powerful source of axions
and the so-called `axion helioscope' technique
may enable us to detect such axions directly
\cite{sikivie1983,bibber1989}.

\begin{figure}
  \includegraphics[scale=0.8]{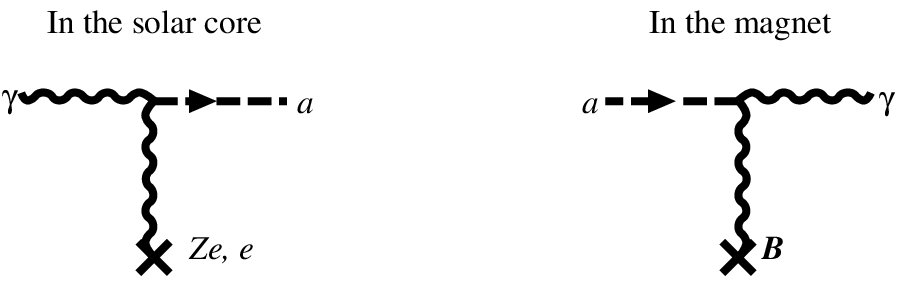}
  \caption{The solar axions produced via the Primakoff process
    in the solar core are, then, converted into X-rays
    via the reverse process in the magnet.}
  \label{fig:principle}
\end{figure}

The principle of the axion helioscope is illustrated
in Fig.~\ref{fig:principle}.
Axions would be produced
through the Primakoff process in the solar core.
The differential flux of solar axions
at the Earth is approximated by
\cite{bahcall2004,raffelt2005}
\begin{eqnarray}
  \d \Phi_a/\d E&=&6.020\times10^{10}[\mathrm{cm^{-2}s^{-1}keV^{-1}}]
  \nonumber\\
  &&{}\times\left(\gagg\over10^{-10}\mathrm{GeV}^{-1}\right)^2
  \left( \frac{E}{1\,\mathrm{keV}}\right)^{2.481}
  \exp \left( -\frac{E}{1.205\,\mathrm{keV}}\right),
  \label{eq:aflux}
\end{eqnarray}
where
$\gagg$ is the axion-photon coupling constant%
\footnote{The formula is different from the one
in our previous paper \cite{sumico1997,sumico2000},
but both coincide numerically with good approximation.}%
.
Their average energy is 4.2\,keV
reflecting actually the core temperature of the sun($\sim 3kT$), 
since low energy axion production is suppressed due to screening effects\cite{raffelt1986}.
Then, they would be coherently converted into X-rays
through the inverse process
in a strong magnetic field at a laboratory.
The conversion rate is given by
\begin{equation}
P_{a\to\gamma} = \left| \frac{g_{a\gamma\gamma}}{2} 
\exp\left[-\int_{0}^{L} \mathrm{d}z \Gamma/2 \right] \times \int_{0}^{L}dz B_\bot\exp
\left[i \int_{0}^{z} \mathrm{d}z^{\prime}\left( q - \frac{i\Gamma}{2}\right) \right]\right|^2,
  \label{eq:prob}
\end{equation}
where
$z$ and $z^{\prime}$ are the coordinate along the incident solar axion,
$B_\bot$ is the strength of the transverse magnetic field,
$L$ is the length of the field along $z$-axis,
$\Gamma$ is the X-ray absorption coefficient of helium,
$q=(m_\gamma^2-m_a^2)/2E$ is the momentum transfer
by the virtual photon,
and 
$m_\gamma$ is the effective mass of the photon
which equals zero in vacuum.
Eq. (\ref{eq:prob}) is reduced to
\begin{equation}
P_{a\to\gamma} = \left( \frac{g_{a\gamma\gamma}B_\bot L}{2}\right)^2
\left[ \frac{\sin(qL/2)}{qL/2}\right]^2,
  \label{eq:prob_plain}
\end{equation}
in case $q$ and $B_\bot$ is constant along $z$-axis and $\Gamma=0$.

In 1997,
the first phase measurement \cite{sumico1997}
was performed
using an axion helioscope with a dedicated superconducting magnet
which is identical to the one used in the present experiment, 
except that the gas container was absent and
the conversion region was vacuum.
Its sensitivity was limited to the axion mass region of $m_a<0.03\rm\,eV$
due to a loss of coherence by non-zero $q$ in
Eq.~(\ref{eq:prob_plain}).

If one can adjust $m_\gamma$ to $m_a$,
coherence will be restored
for non-zero mass axions.
This is achieved by filling the conversion region with gas.
A photon in the X-ray region acquires a positive effective mass
in a medium.
In light gas,
such as hydrogen or helium,
it is well approximated by
\begin{equation}
  m_\gamma=\sqrt{4\pi\alpha N_e\over m_e},
\end{equation}
where $\alpha$ is the fine structure constant,
$m_e$ is the electron mass,
and $N_e$ is the number density of electrons.
We adopted cold helium gas as a dispersion-matching medium.
Here, light gas was preferred since it minimizes self absorption
by gas.
It is worth noting that
helium remains at gas state even at 5--6\,K,
the operating temperature of our magnet.
Since the bore of the magnet is limited in space,
the easiest way is to keep the gas
at the same temperature as the magnet.
Moreover,
axions as heavy as a few electronvolts
can be reached
with helium gas of only about one atmosphere
at this temperature.

In this way,
in 2000,
the second phase measurement \cite{sumico2000}
was performed
to search for sub-electronvolt axions.
This experiment,
together with the first phase measurement of 1997 \cite{sumico1997}
with vacuum conversion region,
yielded an upper limit of
$\gagg<\hbox{6.0--10.5}\times10^{-10}\rm GeV^{-1}$ (95\% CL)
for $m_a<0.27\rm\,eV$.

In this Letter, we will present the result of
the third phase measurement
in which we scanned the mass region between
$0.84<m_a<1.00\rm\,eV$
using the upgraded apparatus to withstand higher pressure gas.

%
%
\section{Experimental apparatus}
\label{sec:app}

\begin{figure}
  \includegraphics[scale=0.8]{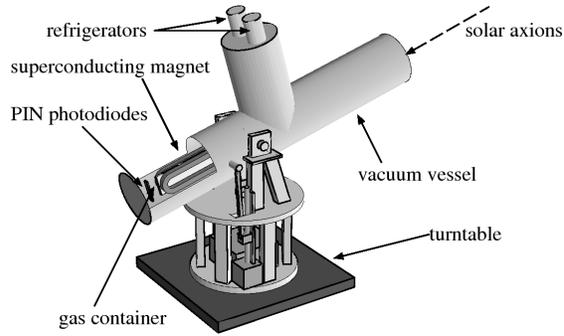}
  \caption{The schematic view of the axion helioscope.}
  \label{fig:sumico}
\end{figure}

The schematic figure of the axion helioscope is shown
in Fig.~\ref{fig:sumico}.
It is designed to track the sun in order to achieve
long exposure time.
It consists of
a superconducting magnet, X-ray detectors, a gas container,
and an altazimuth mounting.
In the following paragraphs,
we will describe each part in due order.

The superconducting magnet \cite{sato1997}
consists of two 2.3-m long
race-track shaped coils running parallel
with a 20-mm wide gap between them.
The magnetic field in the gap is 4\,T
perpendicular to the helioscope axis.
The coils are kept at 5--6\,K during operation.
In order to make it easy to swing this large cryogenic apparatus,
two devices are engaged.
First, the magnet was made cryogen-free
by making two Gifford-McMahon refrigerators
to cool it directly by conduction.
Second, a persistent current switch was equipped.
Thanks to this, the magnet can be freed from
thick current leads after excitation,
and the magnetic field is very stable for a long period of time
without supplying current.

The container to hold dispersion-matching gas is inserted
in the $20\times92\,\mathrm{mm^2}$
aperture of the magnet.
Its body is made of four 2.3-m long 0.8-mm thick
stainless-steel square pipes
welded side by side to each other.
The entire body is wrapped with 5N high purity
aluminium sheet to achieve high uniformity of temperature.
The measured thermal conductance between the both ends was
$1\times10^{-2}\mathrm{W/K}$ at 6\,K.
One end at the forward side of the container is sealed
with welded plugs
and is suspended firmly by three Kevlar cords,
so that thermal flow through this end is highly suppressed.
Combined thermal conductance of three Kevlar cords is estimated to be
smaller than that of the container itself by four orders of magnitude.
Since place-to-place temperature difference of the magnet is likely to be
much less than 1K, the temperature of the container is calculated to be
uniform along the length within the order of 0.1mK, which is fairly
small and has only negligible effect on the coherence of the axion to
photon conversion.  Radiative thermal flow between the gas container and
the magnet is the same order of magnitude as the conductive flow through
the Kevlar cords, and hence negligible, too.
The opposite side nearer to the X-ray detectors
is flanged and fixed to the magnet.
At this end of the container, gas is separated from vacuum
with an X-ray window manufactured by METOREX
which is transparent to X-ray above 2\,keV
and can hold gas up to 0.3\,MPa at liquid helium temperature.

To have automatic sequential pressure settings of the dispersion-matching gas
for the scan of the axion mass region around 1\,eV,
a gas handling system is built with
3 Piezo valves (two HORIBASTEC PV1101 and a PV1302)
and a precision pressure gauge (YOKOGAWA MU101-AH1N).
The temperature of the gas container was measured by a Lakeshore CGR thermistor.

Required pressure for a given mass was determined by the corresponding gas density
and the temperature
based on interpolation of the tables from NIST \cite{nist}.
The error of the target mass is estimated to be less than 5\,meV
by the errors of the pressure and the temperature.
Since we are scanning a range of the axion mass,
the error is only crucial to the lower and upper edges of the mass range.

Helium gas is fed to the container through the Piezo valve to have a specified
pressure setting.
If a lower pressure setting is required, then another Piezo valve is opened
to suck the gas by a vacuum pump connected to the valve.
Once a proper pressure setting is settled, all the valves are closed and the helium gas
is kept confined to have a constant electron number density in the container
until the measurement for the setting is completed.
The uniformity of the temperature guarantees the homogeneous
density along the length of the container.
The whole process is done step by step automatically to scan the axion mass.

Absorption of X-ray in the helium gas is not negligible and the effect is properly
calculated in Eq. (\ref{eq:prob}).
The electron number density might vary slightly along the container
from the one end to the other
because of the gravity
when the inclination is high,
and the coherence could be partly lost accordingly.
The effect is also taken into account in Eq. (\ref{eq:prob}).
The decreases of the conversion probability $P_{a\to \gamma}$
due to the absorption and the gravity are less than 23\% and 1\%
, respectively
when $m_\gamma$ is tuned to 1.0\,eV at the center of the gas container.

For emergency exhaust of the gas in case of rapid temperature increase
due to a magnet quenching,
a rupture disk, which is designed to break at 0.248 MPa,
is introduced into the gas handling system
to avoid destruction of the X-ray window by the over pressure.

Sixteen PIN photodiodes, Hamamatsu Photonics S3590-06-SPL,
are used as the X-ray detectors,
whose chip sizes are $11\times11\times0.5\rm\,mm^3$ each.
In the present measurement, however, twelve of them are used for the analysis
because four went defective through thermal stresses since the measurement of the previous phase.
The effective area of a photodiode was measured
formerly using a pencil-beam X-ray source,
and found to be larger than $9\times9\,\mathrm{mm^2}$.
It has an inactive surface layer of 
$0.35\,\mu\mathrm{m}$ \cite{akimotoPIN}.
Each chip is mounted on a Kapton film
bonded to an Invar plate with cryogenic compatible adhesive.
The X-ray detectors
are mounted in a 10-mm thick radiation shielding box made of
oxygen-free high conductivity copper (OFHC Cu),
which is then surrounded by a lead shield of about 150\,mm thick.
The copper shield is operated at about 60\,K,
so that
it also functions as a cold finger for the X-ray detectors.
Details on the X-ray detector are given
in Refs.\ \cite{naniwaPIN,akimotoPIN}.

The output from each photodiode is fed
to a charge sensitive preamplifier whose first-stage
FET is at the cryogenic stage near the photodiode chip
and the preamplifier outputs are digitized using
CAMAC flash analog-to-digital convertors (FADCs), REPIC RPC-081's,
at a sampling rate of 10 MHz.
The preamplifier outputs are also fed to
shaping amplifiers, Clear Pulse CP4026,
whose outputs are then discriminated to generate triggers.
Thus, waveforms of the sixteen preamplifier outputs
are recorded simultaneously
over 50 $\mu$s before and after each trigger
to be committed to later off-line analysis.
Each detector was calibrated by 5.9-keV Mn X-rays
from a \nuc{55}{Fe} source installed in front of them.
The source is manipulated from the outside
and is completely retracted behind the shield
during the axion observations, i.e. during solar tracking.

The entire axion detector is constructed in a vacuum vessel
and the vessel is mounted on an altazimuth mount.
Its trackable altitude ranges from $-28^\circ$ to $+28^\circ$
and its azimuthal direction is designed to be limited only
by a limiter which prevents the helioscope from endless rotation.
However, in the present measurement, the azimuthal range is restricted to
about 60$^\circ$ because a cable handling system for its
unmanned operation is not completed yet.
The range corresponds to an exposure time
of about a quarter of a day in observing the sun.
This is enough for the time being,
since background is measured during the other three quarters of a day.
When the entire cable handling system is complete, running time per pressure setting
can be shortened by a factor of more than two.

This helioscope mount is driven by two AC servo motors
controlled by a computer (PC).
The PC also monitors the azimuthal and altitudinal directions
of the helioscope regularly
by two precision rotary encoders
and forms a feedback controlling loop as a whole.
The US Naval Observatory Vector Astronomy Subroutines (NOVAS) \cite{novas}
were used to calculate the solar position.
The altitudinal origin was determined from a spirit level.
While the sun is not directly visible from the laboratory
in the basement floor,
the azimuthal origin was first determined
by a gyrocompass, which detects the north direction by the
rotation of the earth within an error of $\pm8"$,
and then it was introduced to the laboratory with a theodolite.

Since the effective aperture of the helioscope is narrow,
it is crucial to determine its accurate geometry.
The axis of the helioscope is defined by
two cross hairs at the edge of the vacuum vessel.
The position of each part of the helioscope was measured
relative to these cross hairs
from their exterior
using the theodolite
when they were installed.
The positions of the PIN photodiodes were determined relative to
the copper shielding box from a photo image
taken prior to the installation.
As it is hard to estimate analytically
the effect of the geometrical errors
as well as the effect of the size of the axion source,
we performed a Monte Carlo simulation and
found that the overall effective area is larger than $371\,\mathrm{mm}^2$
at 99\% confidence level.
It is smaller than the nominal sum of the area of all the living PIN
diodes mainly because the line of sight is partly shaded by the walls of the
four square pipes of the gas container from the axion source region of
the sun.  Also included are many other small effects like geometrical
misalignment of the container and the PIN diodes, distortion of the
square pipes, absorption by thick supporting grids of the X-ray window,
and so on.

%
%
\section{Measurement and Analysis}


From December 2007 through April 2008,
a measurement employing dispersion-matching gas was performed
for 34 photon mass settings with about three days of running time per setting
to scan around 1 eV,
which is shown in Table \ref{tab:settings}.

Before obtaining energy spectra,
each event was categorized into two major groups,
the solar observation and the background.
Events while the measured direction agreed with the sun
are counted as former.
When the sun is completely out of the magnet aperture,
events are counted as latter.
Otherwise events are discarded.

We performed numerical pulse shaping to the raw waveforms
using the Wiener filter.
The energy of an X-ray is given by the peak height of
a wave after the shaping.
The shaped waveform is given by
\begin{equation}
  \label{eq:wiener}
  U(\omega)=
  {S^\ast(\omega) C(\omega)
    \over|N(\omega)|^2},
\end{equation}
where $U(\omega)$, $S(\omega)$, $C(\omega)$,
and $N(\omega)$ are Fourier transformations of
the shaped waveform,
the ideal signal waveform,
the measured waveform,
and the noise, respectively.
Noises are obtained by gathering waveforms
while no trigger exists,
and the ideal signal waveform is
approximated by averaging signals from 5.9-keV X-rays.
The response function of this waveform analysis,
i.e., non-linearity, gain walk by trigger timing, etc.,
was investigated thoroughly using simulated pulses
which were obtained by adding
the template waveform to the noise waveforms.
A correction was made
based on this numerical simulation.
Saturation arose at about 25\,keV,
therefore, $E>20\,\mathrm{keV}$ was not used
in the later analysis.

Event reduction process is applied 
in the same way as the second phase measurement \cite{sumico2000}
in order to get rid of bad events like microphonics.
Applying the same cuts to the $^{55}$Fe source data,
we found the loss of axion detection efficiency by the reduction to be less than 1.5\%.

The background level was
about $1.0\times10^{-5}\mathrm{keV^{-1}s^{-1}/PIN(81\,\mathrm{mm}^2)}$
at $E=5\hbox{--}10\,\mathrm{keV}$.
By analysing the calibration data,
we found the energy resolution of each photodiode
to be
0.8--1.2\,keV (FWHM)
for 5.9-keV photons.

\begin{figure}
  \hbox{%
    \includegraphics[scale=0.5]{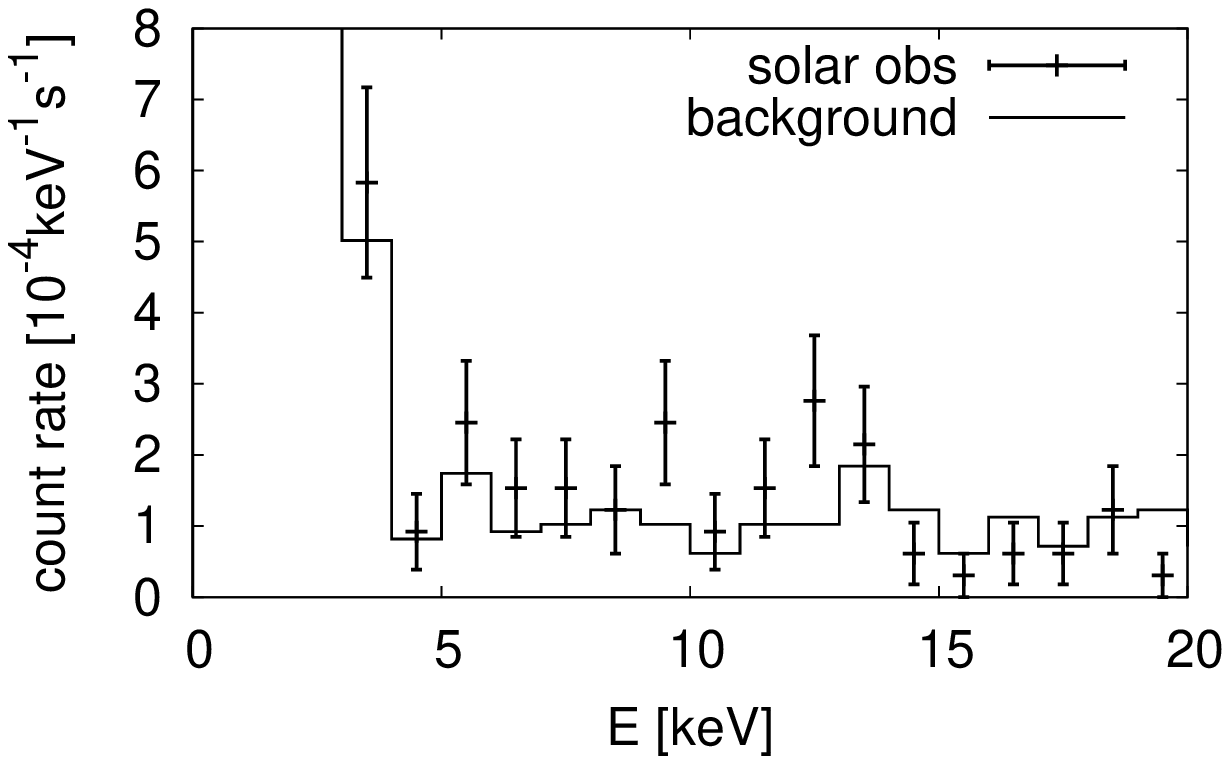}%
    \hskip 1cm
    \includegraphics[scale=0.5]{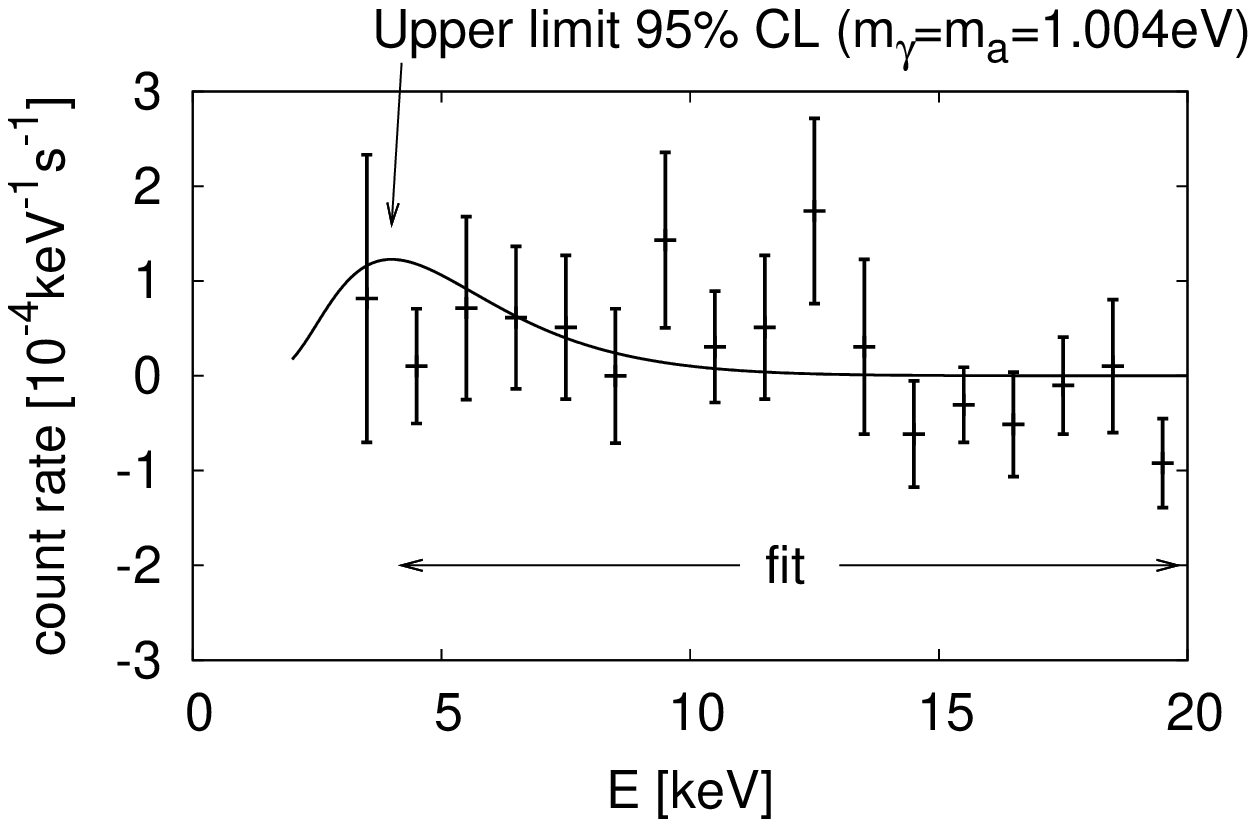}
    }
  \caption{The left figure shows the energy spectrum
    of the solar observation (error bars) and
    the background spectrum (solid line) for the effective PIN photodiode area of 371\, mm$^2$
    when
    the gas density was tuned to $m_\gamma=1.004\rm\,eV$.
    The right figure shows the net energy spectrum of the left
    where the background is subtracted from the solar observation.
    The solid line shows the expected solar axion energy spectrum.
    }
  \label{fig:spec}
\end{figure}

In Fig.~\ref{fig:spec},
the energy spectrum of the solar observation
with the gas density for $m_\gamma=1.004\rm\,eV$
is shown together with the background spectrum.
We searched for expected axion signals
which scale with $\gagg^4$
for various $m_a$
in these spectra.
The smooth curve in the figure represents an example for
the expected axion signal
where $m_a=m_\gamma=1.004\rm\,eV$ 
and $\gagg=7.7\times10^{-10}\rm GeV^{-1}$,
which corresponds to the upper limit
estimated as follows.


A series of least $\chi^2$ fittings was performed assuming various
$m_a$ values.
Data from the 34 different gas density settings
were combined by using the summed $\chi^2$ of the 34.
The energy region of 4--20\,keV was used for fitting
where the efficiency of the trigger system is almost 100\%
and the FADCs do not saturate.
As a result, no significant excess was seen for any $m_a$,
and thus an upper limit on $\gagg$ at 95\% confidence level
was given following the Bayesian scheme.
Fig.~\ref{fig:exclusion} shows
the limit plotted as a function of $m_a$.
Our previous limits from the first \cite{sumico1997}
and the second \cite{sumico2000} phase measurements
and some other bounds are also
plotted in the same figure.
The shown previous limits have been updated using newly measured
inactive surface layer thickness of the PIN photodiode~\cite{akimotoPIN};
the difference is, however, marginal.
The SOLAX~\cite{solax1999}, COSME~\cite{cosme2002} and DAMA~\cite{DAMA2001} are solar axion experiments
which exploit the coherent conversion(i.e. axion Bragg scattering\cite{Pascos})
on the crystalline planes in a germanium and a NaI detector.
The experiment by Lazarus \etal~\cite{Lazarus} and CAST~\cite{CAST} 
are the same kind of experiments as ours.
The latter utilizes a large decommissioned magnet of the LHC at CERN.
Its limit is better than our previous limits by a factor of seven
in low $m_a$ region
due to its larger $B$ and $L$ in Eq.~(\ref{eq:prob_plain}).
In the region $m_a > 0.14\rm\,eV$, however,
our previous and present limits surpass the limit of CAST.
The limit $\gagg<2.3\times10^{-9}\rm GeV^{-1}$ is
the solar limit inferred from the solar age consideration
and the limit $\gagg<1\times10^{-9}\rm GeV^{-1}$
is a more stringent limit reported
by Schlattl \etal ~\cite{schlattl1999}
based on comparison between
the helioseismological sound-speed profile
and
the standard solar evolution models with energy losses by solar axions.
Watanabe and Shibahashi \cite{watanabe2001} have argued
that the helioseismological bound
can be lowered to $\gagg<4.0\times10^{-10}\rm GeV^{-1}$
if the `seismic solar model' and
the observed solar neutrino flux are combined.
%
%
\section{Conclusion}
\begin{figure}
 \hbox{%
  \includegraphics[scale=0.8]{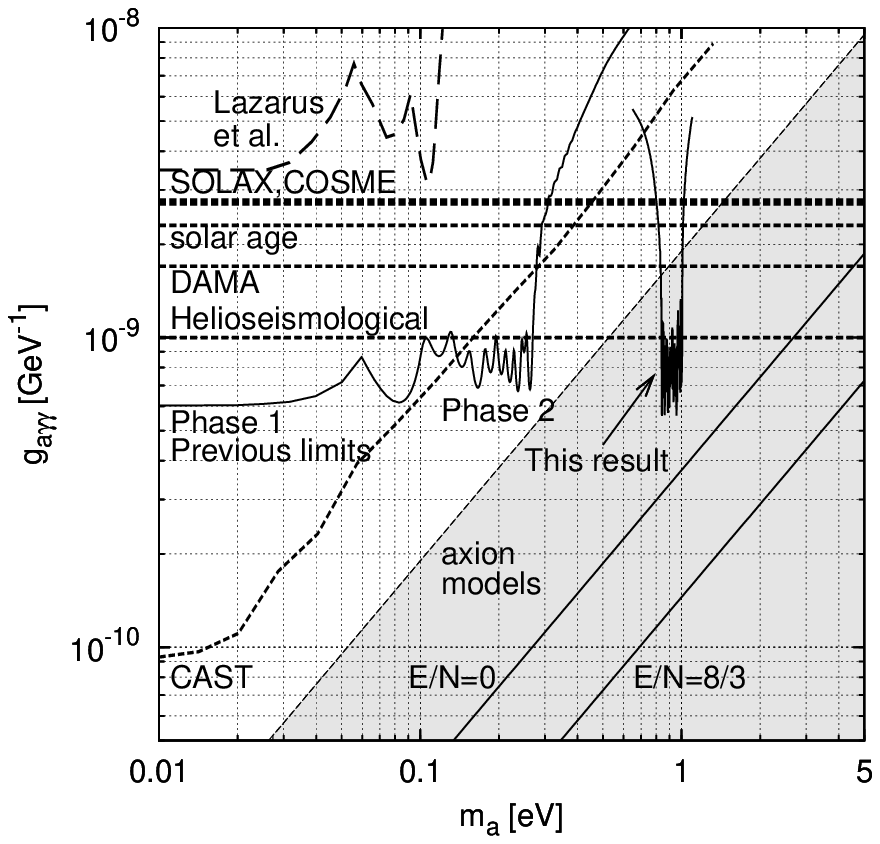}
  \hskip 1cm
  \includegraphics[scale=0.8]{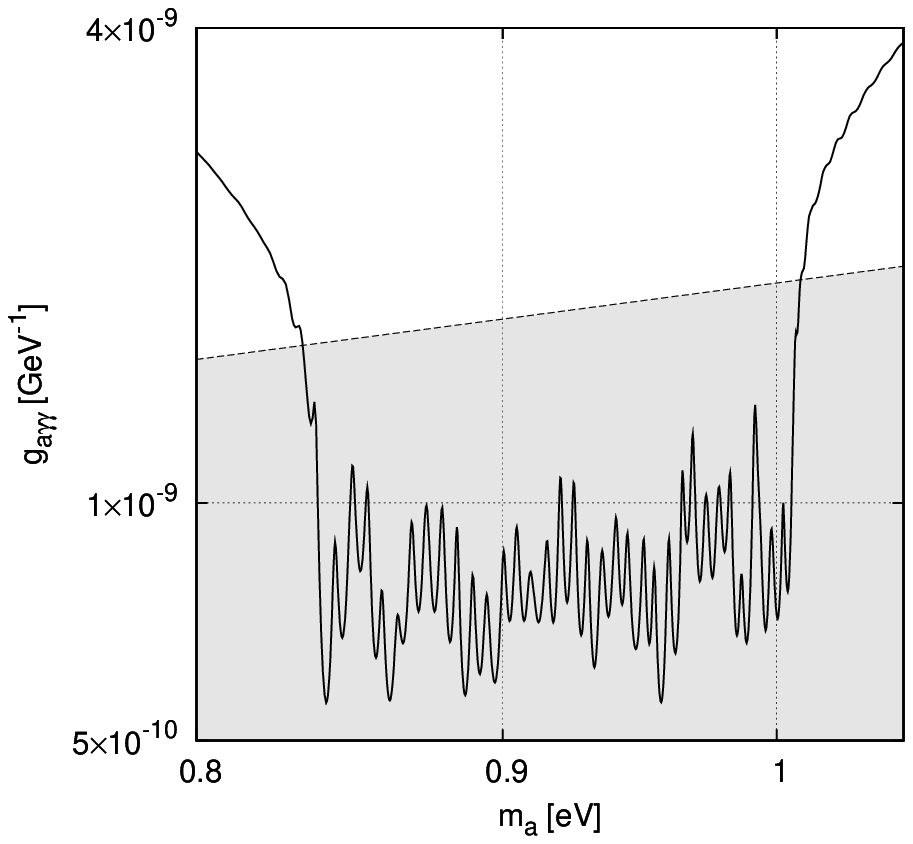}
  }
\caption{The left figure is the exclusion plot on $\gagg$ to $m_a$.
    The new limit and the previous ones\cite{sumico1997,sumico2000} are plotted in solid lines.
    Dashed lines are
    the limit by Lazarus~\etal~\cite{Lazarus},
    the limit by CAST experiment~\cite{CAST},
    the limit by SOLAX experiment~\cite{solax1999},
    the limit by COSME experiment~\cite{cosme2002},
    the limit by DAMA experiment~\cite{DAMA2001},
    the limit inferred from the solar age consideration,
    and the helioseismological bound.
    The hatched area corresponds to the preferred axion models~\cite{GUT_axion}.
    The right figure shows the magnified view of the new limit.
    }
  \label{fig:exclusion}
\end{figure}

The axion mass around 1\,eV has been scanned
with an axion helioscope with cold helium gas
as the dispersion-matching medium
in the $4{\rm\,T}\times2.3\rm\,m$ magnetic field,
but no evidence for solar axions was seen.
A new limit on $\gagg$ shown in Fig. \ref{fig:exclusion}
was set for $0.84<m_a<1.00\rm\,eV$.
It is the first result to search for the axion in the
$\gagg$-$m_a$ parameter region of the preferred axion models~\cite{GUT_axion}
with a magnetic helioscope.

When the complete unmanned operation is ready
with the automatic cable handling system,
the mass scan will be continued to cover still wider mass range around 1\,eV.

\begin{ack}
The authors thank the former director general of KEK, Professor H. Sugawara,
for his support in the beginning of the helioscope experiment.
This research was partially supported
by the Japanese Ministry of Education, Science, Sports and Culture,
Grant-in-Aid for COE Research and Grant-in-Aid for Scientic Research (B),
and also by the Matsuo Foundation.
\end{ack}

%
%

%
%

\clearpage
%
%
\renewcommand{\arraystretch}{1.2}
\begin{table}
  \begin{center}
    \begin{tabular}{ccrr}
      \hline
      molar density&
      $m_\gamma$&
      \multicolumn{2}{c}{live time [s]}\\
      $[\rm mol/m^3]$&
      $[\mathrm{eV}]$&
      \multicolumn{1}{c}{solar run}&
      \multicolumn{1}{c}{background}\\
      \hline
      425.8&0.841&8726&33122\\
431.1&0.846&25362&97561\\
437.1&0.852&25916&97863\\
442.4&0.857&24209&91178\\
447.2&0.862&16724&62828\\
451.7&0.866&23543&92698\\
457.1&0.871&22070&84977\\
462.8&0.877&20242&78517\\
468.3&0.882&20508&79185\\
473.8&0.887&18916&73595\\
479.3&0.892&19569&74852\\
484.8&0.897&18788&72368\\
490.3&0.902&18801&71656\\
495.8&0.907&18985&71630\\
501.4&0.912&18343&68041\\
507.0&0.918&18126&67753\\
512.5&0.923&18022&66218\\
518.1&0.928&18468&68035\\
523.4&0.932&17805&63160\\
529.0&0.937&18683&66060\\
534.6&0.942&19309&65930\\
540.2&0.947&18650&63162\\
545.9&0.952&19801&67513\\
550.9&0.957&20626&68302\\
557.0&0.962&22478&73061\\
562.0&0.966&22126&66594\\
567.6&0.971&24790&74858\\
573.1&0.976&24464&69696\\
578.5&0.980&23120&69297\\
584.0&0.985&23922&73955\\
588.4&0.988&29533&91299\\
597.0&0.996 & 35946&110559\\
602.6&1.000&36503&108174\\
607.3&1.004&32584&97671\\
      \hline
    \end{tabular}
    \caption{Table of the gas settings and each live time.}
    \label{tab:settings}
  \end{center}
\end{table}
\end{document}